\documentclass[]{article}

\usepackage[english]{babel}
\usepackage{amsmath, amsfonts, amssymb, amsthm}
\usepackage{algorithm, algorithmic}
\usepackage[capitalise]{cleveref}
\usepackage{graphicx}

\title{Pell hyperbolas in DLP-based cryptosystems}
\author{Alecci Gessica, Dutto Simone, Nadir Murru}
\date{}

\begin{document}

\maketitle

\begin{abstract}
We present a study on the use of Pell hyperbolas in cryptosystems with security based on the discrete logarithm problem. Specifically, after introducing the group's structure over generalized Pell conics (and also giving the explicit isomorphisms with the classical Pell hyperbolas), we provide a parameterization with both an algebraic and a geometrical approach. The particular parameterization that we propose appears to be useful from a cryptographic point of view because the product that arises over the set of parameters is connected to the Rédei rational functions, which can be evaluated in a fast way. Thus, we exploit these constructions for defining three different public key cryptosystems based on the ElGamal scheme. We show that the use of our parameterization allows to obtain schemes more efficient than the classical ones based on finite fields.
\end{abstract}

\section{Introduction}
\label{sec:intro}

The Pell hyperbola over a field $\mathbb{K}$ is a curve defined for a fixed non-zero element $d \in \mathbb{K}$ as
\begin{equation*}
\mathcal{C}_d(\mathbb{K}) = \big\{ (x, y) \in \mathbb{K} \times \mathbb{K} \,\vert\, x^2 - d y^2 = 1 \big\},
\end{equation*}
whose name comes from the famous Pell equation.
It is well-known that a group structure over the Pell hyperbola can be obtained by considering the Brahmagupta product that, given two points $(x_1, y_1), (x_2, y_2) \in \mathcal{C}_d(\mathbb{K})$, is
\begin{equation}
\label{eq:brah}
(x_1, y_1) \otimes_d (x_2, y_2) = (x_1 x_2 + d y_1 y_2, x_1 y_2 + y_1 x_2).
\end{equation}
The identity element is the vertex of the hyperbola with coordinates $(1, 0)$ and the inverse of a point $(x, y)$ is $(x, - y)$.
When it is clear from the context, we write $\otimes$ instead of $\otimes_d$, omitting the subscript.

If $\mathbb{K}$ is a finite field of order $q$, with $q$ odd prime or power of an odd prime, then the obtained group is cyclic of order $q - \chi(d)$ (see, e.g., \cite{MV92}), where $\chi(d)$ is the quadratic character of $d \in \mathbb{F}_q$, i.e.
\begin{equation*}
\chi(d) = \begin{cases} 
0 & \text{if } d = 0, \\
1 & \text{if } d \text{ is a quadratic residue in } \mathbb F_q, \\
-1 & \text{if } d \text{ is a quadratic non-residue in } \mathbb F_q.
\end{cases}
\end{equation*}

All Pell hyperbolas with same value of $\chi(d)$ are isomorphic.
In particular, if $\chi(d) = \chi(d')$, then $d = d' s^2$ for some $s \in \mathbb{F}_q$ and the isomorphism is given by
\begin{equation}
\label{eq:iso}
\begin{split} 
\big( \mathcal{C}_{d}(\mathbb{F}_q ), \otimes_{d} \big) & \longrightarrow 
\big( \mathcal{C}_{d'}(\mathbb{F}_q), \otimes_{d'} \big), \\
(x, y) & \longmapsto (x, s y),
\end{split}
\end{equation}
because
\begin{equation*}
1 = x^2 - d y^2 = x^2 - d' s^2 y^2 = x^2 - d' (s y)^2.
\end{equation*}

The group's structure of the Pell hyperbola was used in various cryptosystems, especially for constructing RSA-like schemes.
In \cite{Lem06}, an analogue of the RSA scheme over the Pell hyperbola is introduced.
However, it requires to send twice as many bits per message with respect to classic RSA without increasing the security.
For overcoming this issue, other works exploited parameterization of the Pell hyperbola, obtaining RSA-like schemes having a two times faster decryption procedure than RSA \cite{BM16, MS06, Pad06}.
Moreover, recently, several works have exploited these schemes in different contexts \cite{BM20, DRS21, RAB13, TMS20}.

Since the group of the Pell hyperbola is cyclic, we can think to exploit it also in Public-Key Encryption (PKE) schemes whose security is based on the Discrete Logarithm Problem (DLP), such as the ElGamal PKE scheme \cite{ElG85}.
In its classic version, the ElGamal scheme is based on the cyclic multiplicative group of a finite field $\mathbb F_p$, with $p$ prime, and its security is guaranteed by the hardness of the DLP, which can be solved only in sub-exponential time by the index calculus algorithm.
Later, an analogue of the ElGamal scheme over elliptic curves was implemented \cite{Kob87}.
Over the years, the ElGamal scheme was modified in order to speed up the execution time and to increase the efficiency of the scheme \cite{HA19, RA21}.
In \cite{IA18}, authors presented an ElGamal-like cryptosystem based on the matrices over a groupring and, in \cite{Mah08, Mah12}, the authors studied an PKE scheme, called the MOR cryptosystem, similar to the ElGamal scheme over a non-abelian finite group and a group of outer automorphism.
Further studies and variant can be found in \cite{KC14, Kha10, Mah15, PGS20}.

As we will highlight in the next sections, there is a strict link between the group's law of the Pell hyperbola and the elliptic curves, indeed the operation can be geometrically introduced with a a similar construction, but with a more simple algebraic expression in the case of the Pell hyperbola.
Thus, it seems interesting to translate the ElGamal scheme over the Pell hyperbola, exploiting also some specific parameterization of this curve that provide some benefits from a computational point of view.
In particular, in \cref{sec:gen} we will study a generalization of the group's structure on Pell conics defined by the equation $x^2 - d y^2 = c$, writing the group's operation in terms of Brahmagupta products and providing an explicit isomorphisms with the classic Pell hyperbolas.
In \cref{sec:par}, we will provide a parameterization for Pell hyperbolas exploiting an algebraic approach and connecting it also to a geometrical interpretation.
The particular parameterization that we propose appears to be very useful from a cryptographic point of view because the product that arises over the set of parameters is connected to Rédei rational functions, as shown in \cref{sec:exp}.
Then \cref{sec:pke} will be devoted to the presentation of three new PKE schemes based on Pell hyperbolas and in \cref{sec:res} we will present numerical results.

\section{Generalized Pell conics}
\label{sec:gen}

In this section, we introduce a generalization of the group's structure over a generalized Pell conic and we give an explicit isomorphism between a classic Pell hyperbola and a generalized Pell conic.

The equation of the Pell hyperbola is a particular case of the canonical form of hyperbolas and ellipses that, over a finite field, is given by
\begin{equation*} 
\label{eq:gen}
\mathcal{C}_{c, d}(\mathbb{F}_q) = \big\{ (x, y) \in \mathbb{F}_q \times \mathbb{F}_q \,\vert\, x^2 - d y^2 = c \big\}.
\end{equation*}
In addition, the Brahmagupta product can be generalized by considering as identity any point $(a, b) \in \mathcal{C}_{c, d}(\mathbb{F}_q)$.
Thus, we define the group over a generalized Pell conic as $\big( \mathcal{C}_{c, d}(\mathbb{F}_q), \otimes_{a, b, c, d} \big)$, where we denote the generalized Brahmagupta product by $\otimes_{a, b, c, d}$ for highlighting the dependence on the constants $c$ and $d$ that identify the conic, and on the chosen identity point $(a, b)$.
In particular, this product can be obtained from the classic Brahmagupta product $\otimes_d$ as
\begin{equation}
\label{eq:genbrah}
(x_1, y_1) \otimes_{a, b, c, d} (x_2, y_2) = \frac{1}{c} (a, - b) \otimes_d (x_1, y_1) \otimes_d (x_2, y_2).
\end{equation}
Clearly, the identity point for $\otimes_{a, b, c, d}$ is the chosen $(a, b)$, while the inverse of a point $(x, y)$ is the point 
\begin{equation*}
\frac{1}{c} (a, b) \otimes_d (a, b) \otimes_d (x, y).
\end{equation*}
When $c = 1$ and the chosen identity point is $(a, b) = (1, 0)$, the product $\otimes_{a, b, c, d}$ coincides with the classic Brahmagupta product $\otimes_d$ from \cref{eq:brah}.
When it is clear from the context we will just write $\otimes_{a, b}$, omitting the subscripts $c$ and $d$.

The order of the obtained group is still $q - \chi(d)$ since:
\begin{itemize}
\item if $\chi(c) = 1$, then there is a bijection with the classic Pell hyperbola
\begin{equation*}
\begin{split}
\mathcal{C}_{d}(\mathbb{F}_q) & \longrightarrow \mathcal{C}_{c, d}(\mathbb{F}_q), \\
(x, y) & \longmapsto (\sqrt{c} x, \sqrt{c} y);
\end{split}
\end{equation*}
\item if $\chi(c) = - 1$, then after fixing $(x_c, y_c) \in \mathcal{C}_{c, d}(\mathbb{F}_q)$ the bijection with the classic Pell hyperbola is
\begin{equation*}
\begin{split}
\mathcal{C}_{d}(\mathbb{F}_q) & \longrightarrow \mathcal{C}_{c, d}(\mathbb{F}_q), \\
(x, y) & \longmapsto (x, y) \otimes_d (x_c, y_c).
\end{split}
\end{equation*}
\end{itemize}

Despite these bijections are not morphisms, we found the explicit isomorphism between two generalized Pell conics with same $d$ by exploiting the definition of the generalized Brahmagupta product $\otimes_{a, b, c, d}$.

Starting from the classic Pell hyperbola $\mathcal{C}_d(\mathbb{F}_q)$ with the classic Brahmagupta product $\otimes_d$, the points of a generalized Pell conic $\mathcal{C}_{c, d}(\mathbb{F}_q)$ with identity $(a, b)$ are obtained through
\begin{equation*}
\begin{split}
\varphi^{a, b}_{c, d}: 
\big( \mathcal{C}_{d}(\mathbb{F}_q), \otimes_d \big) 
& \longrightarrow 
\big( \mathcal{C}_{c, d}(\mathbb{F}_q), \otimes_{a, b, c, d} \big), \\
(x, y) & \longmapsto (a, b) \otimes_d (x, y).
\end{split}
\end{equation*}
This is a morphism since
\begin{equation*}
\begin{split}
\varphi^{a, b}_{c, d} \big( (x_1, y_1) \otimes_d (x_2, y_2) \big) 
& = (a, b) \otimes_d (x_1, y_1) \otimes_d (x_2, y_2) \\
& = \frac{1}{c} (a^2 - d b^2, 0) \otimes_d (x_1, y_1) \otimes_d (a, b) \otimes_d (x_2, y_2) \\
& = \varphi^{a, b}_{c, d}(x_1, y_1) \otimes_{a, b, c, d} \varphi^{a, b}_{c, d}(x_2, y_2),
\end{split}
\end{equation*}
and, since the two conics have same cardinality, it is an isomorphism with inverse
\begin{equation*}
\begin{split}
(\varphi^{a, b}_{c, d})^{-1}: 
\big( \mathcal{C}_{c, d}(\mathbb{F}_q), \otimes_{a, b, c, d} \big) 
& \longrightarrow 
\big( \mathcal{C}_{d}(\mathbb{F}_q), \otimes_d \big), \\
(x, y) & \longmapsto (1, 0) \otimes_{a, b, c, d} (x, y) 
\end{split}
\end{equation*}

We obtained the explicit isomorphism between two generalized Pell conics with same $d$ by applying $(\varphi^{a, b}_{c, d})^{-1}$ and $\varphi^{a', b'}_{c', d}$, resulting in
\begin{equation}
\label{eq:geniso}
\begin{split}
\big( \mathcal{C}_{c, d}(\mathbb{F}_q), \otimes_{a, b, c, d} \big) 
& \longrightarrow 
\big( \mathcal{C}_{c', d}(\mathbb{F}_q), \otimes_{a', b', c', d} \big), \\
(x, y) & \longmapsto 
(a', b') \otimes_{a, b, c, d} (x, y).
\end{split}
\end{equation}

By combining this procedure with the explicit isomorphism between Pell hyperbolas with same quadratic character of $d$ introduced in \cref{eq:iso}, we obtained an explicit isomorphism between all generalized Pell conics with different $a, b, c, d$ but same $\chi(d)$.

From a computational point of view, it is not useful to use a generalized Pell conic in a DLP-based cryptosystem since the security would remain unchanged while the computational costs would arise.
However, in \cref{sec:pke}, we will exploit the isomorphism between Pell hyperbolas with different $d$ but same $\chi(d)$ to obtain an alternative PKE scheme based on the ElGamal scheme.

\section{Parameterization}
\label{sec:par}

In this section, we describe and study a parameterization for Pell hyperbolas with both an algebraic and a geometrical interpretation.
The algebraic construction of the Pell hyperbola allows us to obtain a useful parameterization for any generalized Pell conic.

When considering the quotient ring (or field if $d$ is not a square)
\begin{equation*}
\mathbb{P}_{d, q} = \mathbb{F}_q[t] / (t^2 - d) = \big\{ x + t y \,\vert\, x, y \in \mathbb{F}_q,\, t^2 = d \big\},
\end{equation*}
for any two elements $x_1 + t y_1, x_2 + t y_2 \in \mathbb{P}_{d, q}$, the product naturally induced from the quotient is
\begin{equation*}
(x_1 + t y_1) (x_2 + t y_2) = (x_1 x_2 + d y_1 y_2) + t (x_1 y_2 + y_1 x_2),
\end{equation*}
which is essentially the classic Brahmagupta product, while the Pell hyperbola coincides with the elements of norm $x^2 - d y^2$ equal to one.

Now we can consider the quotient group $\mathcal{P}_{d, q} = \mathbb{P}_{d, q}^\times / \mathbb{F}_q^\times$ that contains the equivalence classes of the form $[x + t y] = \big\{ \lambda (x + t y) \,\vert\, \lambda \in \mathbb{F}_q^\times \big\}$.
Since when $y = 0$ the class is $[x] = [1_{\mathbb{F}_q}]$, we can describe $\mathcal{P}_{d, q}$ in the following way:
\begin{itemize}
\item $\mathcal{P}_{d, q} = \big\{ [m + t] \,\vert\, m \in \mathbb{F}_q \big\} \cup \big\{ [1_{\mathbb{F}_q}] \big\} \cong \mathbb{F}_q \cup \{\alpha\}$, when $d$ is not a square and $\alpha$ not in $\mathbb F_q$;
\item $\mathcal{P}_{d, q} = \big\{ [m + t] \,\vert\, m \in \mathbb{F}_q \smallsetminus \{ \pm \sqrt{d}\} \big\} \cup \big\{ [1_{\mathbb{F}_q}] \big\} \cong \big( \mathbb{F}_q \smallsetminus \{ \pm \sqrt{d}\} \big) \cup \{\alpha\}$, when $d$ is a square.
\end{itemize}
The product induced on $\mathcal{P}_{d, q}$ is given by
\begin{equation*}
[m_1 + t] \odot_d [m_2 + t] = [m_1 m_2 + d + t (m_1 + m_2)],
\end{equation*}
and, if $m_1 + m_2 \not= 0$, we have
\begin{equation*}
[m_1 + t] \odot_d [m_2 + t] = \left[ \frac{m_1m_2 + d}{m_1 + m_2} + t \right],
\end{equation*}
else
\begin{equation*}
[m_1 + t] \odot_d [m_2 + t] = [m_1m_2 + d] = [1_{\mathbb F_q}].
\end{equation*}
Considering the above isomorphic structures, we can also write that
\begin{equation}
\label{eq:parprod}
m_1 \odot_d m_2 = \begin{cases}
\frac{m_1 m_2 + d}{m_1 + m_2} & \text{if } m_1 + m_2 \neq 0, \\
\alpha & \text{otherwise.}
\end{cases}
\end{equation}
When it is clear from the context, we will write $\odot$ instead of $\odot_d$.

The obtained group is isomorphic to the Pell hyperbola through
\begin{equation*}
\begin{split}
\pi_d^{-1}: \big( \mathcal{P}_{d, q}, \odot_d \big) & \longrightarrow \big( \mathcal{C}_d(\mathbb{F}_q,) \otimes_d \big), \\
m & \longmapsto \left( \frac{m^2 + d}{m^2 - d}, \frac{2 m}{m^2 - d} \right), \\
\alpha & \longmapsto (1, 0),
\end{split}
\end{equation*}
since
\begin{equation*}
\begin{split}
\pi_d^{-1} \left(m_1 \odot_d m_2 \right)_x & = \frac{(m_1 m_2 + d)^2 + d (m_1 + m_2)^2}{(m_1 m_2 + d)^2 - d (m_1 + m_2)^2} \\
& = \frac{(m_1^2 + d) (m_2^2 + d) + 4 d m_1 m_2}{(m_1^2 - d) (m_2^2 - d)} \\
& = \frac{m_1^2 + d}{m_1^2 - d} \frac{m_2^2 + d}{m_2^2 - d} + d \frac{2 m_1}{m_1^2 - d} \frac{2 m_2}{m_2^2 - d} \\
& = \left( \pi_d^{-1} (m_1) \otimes_d \pi_d^{-1} (m_2) \right)_x, \\
\pi_d^{-1} \left(m_1 \odot_d m_2 \right)_y & = \frac{2 (m_1 m_2 + d) (m_1 + m_2)}{(m_1 m_2 + d)^2 - d (m_1 + m_2)^2} \\
& = \frac{2 m_1 (m_2^2 + d) + 2 m_2 (m_1^2 + d)}{(m_1^2 - d) (m_2^2 - d)} \\
& = \frac{2 m_1}{m_1^2 - d} \frac{m_2^2 + d}{m_2^2 - d} + \frac{m_1^2 + d}{m_1^2 - d} \frac{2 m_2}{m_2^2 - d} \\
& = \left( \pi_d^{-1} (m_1) \otimes_d \pi_d^{-1} (m_2) \right)_y.
\end{split}
\end{equation*}
The inverse of this isomorphism is a parameterization of the Pell hyperbola
\begin{equation*}
\begin{split}
\pi_d: \big( \mathcal{C}_d(\mathbb{F}_q,) \otimes_d \big) & \longrightarrow \big( \mathcal{P}_{d, q}, \odot_d \big), \\
(x, y) & \longmapsto
\begin{cases}
\frac{x + 1}{y} & \text{if } y \neq 0, \\
0 & \text{if } (x, y) = (- 1, 0), \\
\alpha  & \text{if } (x, y) = (1, 0).
\end{cases}
\end{split}
\end{equation*}
Thus, the algebraic construction of $\mathcal{P}_{d, q}$ defines the parameterization of the Pell hyperbola that can be obtained considering the lines
\begin{equation*}
y = \frac{1}{m}(x + 1),
\end{equation*}
for $m$ varying in $\mathbb F_q$ or $m = \alpha$ (having the sense of the point at the infinity).
It is noteworthy that the definition of $\big( \mathcal{P}_{d, q}, \odot_d \big)$ is independent of the choice of the identity point $(a, b)$ and of the constant $c$, so that all the previous constructions can be adapted for generalized Pell conics, leading to the parameterization
\begin{equation} 
\label{eq:par}
\begin{split}
\pi^{a, b}_{c, d} : \big( \mathcal{C}_{c, d}(\mathbb{F}_q), \otimes_{a, b, c, d} \big) & \longrightarrow \big( \mathcal{P}_{d, q}, \odot_d \big), \\
(x, y) & \longmapsto
\begin{cases}
\frac{x + a}{y - b} & \text{if } y \neq b, \\
-\frac{b d}{a} & \text{if } (x, y) = (- a, b), \\
\alpha  & \text{if } (x, y) = (a, b),
\end{cases}
\end{split}
\end{equation}
with inverse
\begin{equation*}
\begin{split}
(\pi^{a, b}_{c, d})^{- 1} : \big( \mathcal{P}_{d, q}, \odot_d \big) & \longrightarrow \big( \mathcal{C}_{c, d}(\mathbb{F}_q), \otimes_{a, b, c, d} \big), \\
m & \longmapsto
\begin{cases}
\left( 2 m \frac{a m + b d}{m^2 - d} - a,
2 \frac{a m + b d}{m^2 - d} + b \right) & \text{if } m \neq \alpha, \\
(a, b) & \text{otherwise.}
\end{cases}
\end{split}
\end{equation*}
This parameterization and its inverse can be used as an alternative way to obtain the isomorphism in \cref{eq:geniso}.

\begin{figure}[t]
\begin{center}
\includegraphics[scale=.35]{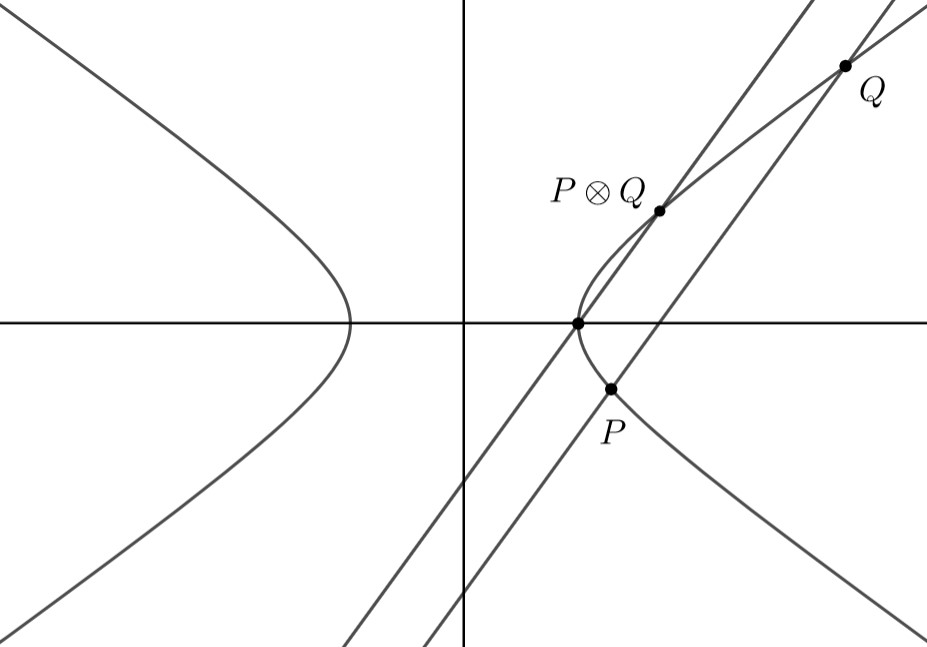}
\end{center}
\caption{Geometric interpretation of the Brahmagupta product.}
\label{fig:prod}
\end{figure}

From a geometrical point of view, the parameter $m$ of a point $(x, y)$ is the slope of the line through $(x, y)$ and $(- a, b)$ written considering $x$ variable with $y$.
This is very interesting when related to the geometric interpretation of the Brahmagupta product, which can be introduced in a very similar way to the one of the elliptic curves.
Indeed, given two points $P$ and $Q$ of an elliptic curve, their sum $P \oplus Q$ is obtained by considering the point $R$, intersection between the elliptic curve and the line through $P$ and $Q$, so that $P \oplus Q$ is the intersection between the elliptic curve and the line through $R$ and the identity point, that is the point at infinity.
This construction works also considering two points $P$ and $Q$ of the Pell hyperbola, with the difference that the line through $P$ and $Q$ intersects the conic at the point $R$ that is, in this case, the point at infinity.
Consequently, the product $P \otimes Q$ is the intersection between the conic and the line through $R$ (point at infinity) and the identity, that is, in the case of the classic Brahmagupta product, the point $(1, 0)$, i.e., the line parallel to the line through $P$ and $Q$ (see \cref{fig:prod}).

It is quite easy to check that, from this geometrical construction of $\otimes$, we obtain the algebraic expression described in \cref{eq:brah}.
Indeed, given $P = (x_1, y_1)$ and $Q = (x_2, y_2)$ on the conic, it is sufficient to check that the slope of the line through $P$ and $Q$ is equal to that of the line through $(x_1 x_2 + d y_1 y_2, x_1 y_2 + y_1 x_2)$ and $(1, 0)$.
Thanks to this geometrical approach, we can also observe that the identity point can be an arbitrary point of the Pell hyperbola, and this choice leads to the generalized Brahmagupta product in \cref{eq:genbrah}.

\section{Exponentiation with Rédei rational functions}
\label{sec:exp}

In this section, we show an efficient algorithm for the exponentiation on the Pell hyperbola that exploits the parameterization introduced in the previous one.
Moreover, we compare it with the square-multiply algorithm with the classic Brahmagupta product.

When dealing with cryptosystems whose security is based on the DLP, the computational bottleneck is the evaluation of the exponentiation, which is usually implemented with a square-multiply algorithm, eventually enhanced with a pre-computation phase.
Therefore the total time is determined by the speed of the single product, which is required in both square and multiply steps.

\begin{figure}
\begin{center}
\begin{minipage}[t]{.45\textwidth}
$\texttt{Brahmagupta}(x_P, y_P, e, d, q)$:

\begin{algorithmic}[1]
\STATE {$(x, y) = (1, 0)$}
\STATE {$\textit{bin\_e} = \texttt{binary}(e)$}
\FOR {$\textit{bit} \textbf{ in } \textit{bin\_e}$}
\STATE {$x = x^2 + d y^2 \in \mathbb{F}_q$}
\STATE {$y = 2 x y \in \mathbb{F}_q$}
\IF {$\textit{bit} = 1$}
\STATE {$x = x x_P + d y y_P \in \mathbb{F}_q$}
\STATE {$y = x y_P + y x_P \in \mathbb{F}_q$}
\ENDIF
\ENDFOR
\RETURN {$(x, y)$}
\end{algorithmic}
\end{minipage}
\begin{minipage}[t]{.5\textwidth}
$\texttt{Modified\_More}(m, e, d, q)$:

\begin{algorithmic}[1]
\STATE {$N, D = 1, 0$}
\STATE {$\textit{bin\_e} = \texttt{binary}(e)$}
\FOR {$\textit{bit} \textbf{ in } \textit{bin\_e}$}
\STATE {$N = N^2 + d D^2 \in \mathbb{F}_q$}
\STATE {$D = 2 N D \in \mathbb{F}_q$}
\IF {$\textit{bit} = 1$}
\STATE {$N = N m + d D \in \mathbb{F}_q$}
\STATE {$D = N + D m \in \mathbb{F}_q$}
\ENDIF
\ENDFOR
\RETURN {$N / D \in \mathbb{F}_q$}
\end{algorithmic}
\end{minipage}  
\end{center}
\caption{Square-multiply algorithm with the Brahmagupta product (left) and modified More's algorithm for Rédei polynomials (right).}
\label{fig:sqmult}
\end{figure}

The Brahmagupta product described explicitly in \eqref{eq:brah} requires 5 products and 2 additions, while the product on the parameters obtained in \eqref{eq:parprod} requires 1 inversion, 2 products and 2 additions.
The inversion is largely more expensive than the additional 3 products required in the Brahmagupta product.
Therefore, in a comparison of square-multiply implementations, the first one is the most efficient.
However, when $d$ is a quadratic non-residue, the product \eqref{eq:parprod} can be evaluated exploiting the Rédei rational functions.
They are introduced over the real numbers by means of the powers
\begin{equation*}
(m + \sqrt{d})^n = A_n(m,d) + B_n(m, d) \sqrt{d},
\end{equation*}
which define two sequences of polynomials whose ratios provide the Rédei rational functions:
\begin{equation*}
Q_n(m, d) = \frac{A_n(m, d)}{B_n(m, d)},
\end{equation*}
with $m, d \in \mathbb{Z} \smallsetminus \{0\}$ and $d$ non-square.
Observing that
\begin{equation*}
Q_1(m, d) = m \quad \text{and} \quad Q_{n_1 + n_2}(m, d) = Q_{n_1}(m, d) \odot_d Q_{n_2}(m, d),
\end{equation*}
we have
\begin{equation*}
m^{\odot e} = Q_e(m, d).
\end{equation*}
Clearly, the above definition can be easily adapted over finite fields.
In \cite{Mor95}, the author proposed an algorithm for evaluating the Rédei rational function $Q_n(m, d)$ with complexity $O(\log n)$ considering additions and multiplications over a ring, and in \cite{BMD21}, the authors improved the performance of this algorithm.
The obtained algorithm is detailed in \cref{fig:sqmult}, where it is compared with the square-multiply algorithm with the classic Brahmagupta product.
In particular, the two algorithms share the quantity of operations at each step except for steps 7/8, where the Brahmagupta version requires an additional product, and step 11, where the modified More's algorithm requires a final inversion.
 
Thus, from an efficiency point of view, the two algorithms are comparable.
The main advantage in adopting the parameterization and the modifies More's algorithm is that the size of the data is halved because they are elements of $\mathbb{F}_q$ and not of $\mathcal{C}_d(\mathbb{F}_q)$.

\section{Public-key encryption with the Pell hyperbola}
\label{sec:pke}

In this section, we present and compare three different PKE schemes based on the ElGamal scheme with Pell hyperbolas over a finite field $\mathbb F_q$.
In particular, we exploit the group on the Pell hyperbola $\mathcal{C}_d(\mathbb F_q)$ with $d$ non square and the Brahmagupta product with $(1, 0)$ as identity, as well as the parameterization presented in \cref{sec:par}.
In the following, it is useful to take $q = p$ prime and $p = 2 p' - 1$ with $p'$ prime in order to avoid small subgroups.

The first algorithm is detailed in \cref{fig:brah}.
Given $q$ power of a prime $n$ bits long (step 1), the order of the cyclic group is $q - \chi(d) = q + 1$.
After choosing, in step 2, $d \in \mathbb{F}_q$, a random generator $G$ of $\mathcal{C}_d(\mathbb{F}_q)$ is taken in step 3.
Then the algorithm proceeds by taking a random exponent $sk$ (step 4) and obtaining a public point $H \in \mathcal{C}_d(\mathbb{F}_q)$ through the square-multiply algorithm with the Brahmagupta product (step 5).
The public key contains $q, d$ and the points $G$ and $H$, while the secret key is the exponent $sk$ used to obtain $H$ from $G$.
In the first step of the encryption algorithm, the message determines the $y$ coordinate of a point, while in the second step the corresponding $x$ is chosen under the condition $(x, y) \in \mathcal{C}_d(\mathbb{F}_q)$.
Since such a point could not exist, some bits of $y$ can be kept variable by reducing the maximum length of the message.
The ciphertext consists of two points $C_1$ and $C_2$.
After taking a random exponent $r$ in step 3, it is used in step 4 to obtain $C_1$ through the exponentiation with the Brahmagupta product with base the public generator $G$.
In step 5, the point $C_2$ is determined as the Brahmagupta product of $H^{\otimes r}$ with the point $(x, y)$ representing the message.
During the decryption, the point $(x, y)$ is retrieved as the Brahmagupta product of the inverse of $C_1^{\otimes sk}$ with $C_2$ (step 1).
From the obtained $y$ coordinate, the original message is recovered in step 2.
In particular, this implementation is formally equivalent to the ElGamal PKE scheme with a cyclic subgroup of order $q + 1$ of $\mathbb{F}_{q^2}$.

\begin{figure}
\begin{center}
\begin{minipage}[t]{.48\textwidth}
$\texttt{KeyGen}(n)$:

\begin{algorithmic}[1]
\STATE {$q \leftarrow_\$ \{0, 1\}^n$ power of a prime}
\STATE {$d \leftarrow_\$ \mathbb{F}_q$ with $\chi(d) = - 1$}
\STATE {$G \leftarrow_\$ \mathcal{C}_d(\mathbb{F}_q)$ of order $q + 1$}
\STATE {$sk \leftarrow_\$ \{2, \ldots, q\}$}
\STATE {$H = G^{\otimes sk} \in \mathcal{C}_d(\mathbb{F}_q)$}
\STATE {$pk = (q, d, G, H)$}
\RETURN {$pk, sk$}
\end{algorithmic}
\end{minipage}  
\begin{minipage}[t]{.48\textwidth}
$\texttt{Encrypt}(\textit{msg}, pk)$:

\begin{algorithmic}[1]
\REQUIRE {$\textit{msg} < q$}
\STATE {$y \leftarrow msg$}
\STATE {$x = \sqrt{1 + d \, y^2} \in \mathbb{F}_q$}
\STATE {$r \leftarrow_\$ \{2, \ldots, q\}$}
\STATE {$C_1 = G^{\otimes r} \in \mathcal{C}_d(\mathbb{F}_q)$}
\STATE {$C_2 = H^{\otimes r} \otimes (x, y) \in \mathcal{C}_d(\mathbb{F}_q)$}
\RETURN {$C_1, C_2$}
\end{algorithmic}

$\texttt{Decrypt}(c_1, c_2, pk, sk)$:

\begin{algorithmic}[1]
\STATE {$(x, y) = (C_1^{\otimes sk})^{-1} \otimes C_2 \in \mathcal{C}_d(\mathbb{F}_q)$}
\STATE {$\textit{msg} \leftarrow y$}
\RETURN {\textit{msg}}
\end{algorithmic}
\end{minipage}
\end{center}
\caption{ElGamal PKE scheme with the cyclic group $(\mathcal{C}_d(\mathbb{Z}_q), \otimes)$ of order $q + 1$.
Exponentiations are realized with the square-multiply algorithm with the Brahmagupta product.}
\label{fig:brah}
\end{figure}

The second algorithm is described in \cref{fig:elg} and consists of the ElGamal PKE scheme with the cyclic group $(\mathcal{P}_{d, q}, \odot)$.
A power of a prime $q$ of $n$ bits is taken in step 1 and the order of the cyclic group is still $q - \chi(d) = q + 1$.
In step 2, a random non-square $d \in \mathbb{F}_q$ is taken.
After choosing a generator $g \in \mathcal{P}_{d, q}$ in step 3 and a random exponent $sk$ in step 4, a parameter $h = g^{\odot sk}$ is evaluated in step 5 with the modified More's algorithm.
The public key consists of $q, d$ and the parameters $g, h$, while the secret key is the exponent $sk$.
The encryption considers the message as a parameter $\textit{msg} \in \mathcal{P}_{d, q}$.
Step 1 takes a random exponent $r$, which is used in step 2 to obtain the parameter $c_1$ through the modified More's algorithm for the exponentiation.
The second parameter $c_2$ is the result of the parameter product between $h^{\odot r}$ and \textit{msg}.
Finally, the ciphertext is the pair of parameters $(c_1, c_2)$ and consequently it requires half of the space than in the previous algorithm.
The decryption is straightforward.
It retrieves the message as the parameter product between the inverse of $c_1^{\odot sk}$ (which is simply its opposite) and $c_2$.
Because of the comparison between the exponentiation algorithms in \cref{sec:exp}, the computational time is comparable with that of the ElGamal scheme with the points of the Pell hyperbola.
However, the public key and the ciphertext require less space because they contain parameters instead of coordinates.

\begin{figure}
\begin{center}
\begin{minipage}[t]{.48\textwidth}
$\texttt{KeyGen}(n)$:

\begin{algorithmic}[1]
\STATE {$q \leftarrow_\$ \{0, 1\}^n$ power of a prime}
\STATE {$d \leftarrow_\$ \mathbb{F}_q$ with $\chi(d) = - 1$}
\STATE {$g \leftarrow_\$ \mathcal{P}_{d, q}$ of order $q + 1$}
\STATE {$sk \leftarrow_\$ \{2, \ldots, q\}$}
\STATE {$h = g^{\odot sk} \in \mathcal{P}_{d, q}$}
\STATE {$pk = (q, d, g, h)$}
\RETURN {$pk, sk$}
\end{algorithmic}
\end{minipage}  
\begin{minipage}[t]{.48\textwidth}
$\texttt{Encrypt}(\textit{msg}, pk)$:

\begin{algorithmic}[1]
\REQUIRE {$\textit{msg} < q$}
\STATE {$r \leftarrow_\$ \{2, \ldots, q\}$}
\STATE {$c_1 = g^{\odot r} \in \mathcal{P}_{d, q}$}
\STATE {$c_2 = h^{\odot r} \odot \textit{msg} \in \mathcal{P}_{d, q}$}
\RETURN {$c_1, c_2$}
\end{algorithmic}

$\texttt{Decrypt}(c_1, c_2, pk, sk)$:

\begin{algorithmic}[1]
\STATE {$\textit{msg} = - c_1^{\odot sk} \odot c_2 \in \mathcal{P}_{d, q}$}
\RETURN {\textit{msg}}
\end{algorithmic}
\end{minipage}
\end{center}
\caption{ElGamal PKE scheme with the cyclic group $(\mathcal{P}_{d, q}, \odot)$ of order $q + 1$ of the parameters of $\mathcal{C}_d(\mathbb{F}_q)$.
Exponentiations are realized with the modified More's algorithm.}
\label{fig:elg}
\end{figure}

The third algorithm is an alternative version of ElGamal PKE scheme with the use of the parameters.
The differences are due to the exploitation of the explicit isomorphisms between Pell hyperbolas with different $d$.
The algorithms are described in \cref{fig:elg2}.
The key generation is analogous to the previous one, except for the smallest non-square $d'$ taken in step 3, which is used for the exponentiation in step 6 and then included in the public key.
The main differences are in the encryption: the maximum length of the message can be doubled with respect to the previous algorithms, because it is used in step 1 to obtain the coordinates of a point $(x, y) \in \mathbb{F}_q \times \mathbb{F}_q$.
From this point, step 2 searches for a quadratic non-residue $d \in \mathbb{F}_q$ such that $(x, y) \in \mathcal{C}_d(\mathbb{F}_q)$.
If necessary, some of the bits of $x$ can be kept variable so that such a $d$ can be found.
Then, in step 3, the parameter $m$ related to the point is obtained through the parameterization from \cref{eq:par}.
In step 4 a random exponent $r$ is chosen.
Now, since the public key contains parameters of points of $\mathcal{C}_{d'}(\mathbb{F}_q)$, the isomorphism between Pell hyperbolas is required: supposing $d = d' s^2$, its explicit formula is
\begin{equation}
\begin{split}
(\mathcal{C}_{d'}(\mathbb{F}_p), \otimes_{d'}) & \longrightarrow (\mathcal{C}_d(\mathbb{F}_p), \otimes_d), \\
(x, y) & \longmapsto (x, s^{- 1} y).
\end{split}
\end{equation}
This gives the isomorphism on the parameters
\begin{equation}
\begin{split}
\big( \mathcal{P}(\mathbb{F}_p), \odot_{d'} \big) & \longrightarrow 
\big( \mathcal{P}(\mathbb{F}_p), \odot_d \big), \\
m = \frac{x + 1}{y} & \longmapsto s m,
\end{split}
\end{equation}
which is used to obtain $g s$ and $h s$ to be used for the exponentiations with $\odot_d$.
In step 5, the factor $s = \sqrt{d / d'} \in \mathbb{F}_q$ can always be evaluated because $d$ and $1 / d'$ are quadratic non-residues and their product is a quadratic residue.
Steps 6-7 evaluate the parameters $c_1, c_2$ as in the previous algorithm but with the basis obtained from the isomorphism with the factor $s$.
The ciphertext contains $c_1, c_2$ and also the quadratic non-residue $d$ used in the calculations.
The setting of the decryption is analogous to the previous case but, after evaluating the product between the inverse of $c_1^{\odot_d sk}$ and $c_2$ (step 1), the message must be retrieved from the point related to the obtained parameter (step 2).
The main advantage in adopting this algorithm is the doubled length of the message with a cost of less than double operations and also the decreased keys and ciphertext sizes.

\begin{figure}
\begin{center}
\begin{minipage}[t]{.48\textwidth}
$\texttt{KeyGen}(n)$:

\begin{algorithmic}[1]
\STATE {$q \leftarrow_\$ \{0, 1\}^n$ power of a prime}
\STATE {$d' \in \mathbb{F}_q$ minimum with $\chi(d')\!=\!- 1$}
\STATE {$g \leftarrow_\$ \mathcal{P}_{d', q}$ of order $q + 1$}
\STATE {$sk \leftarrow_\$ \{2, \ldots, q\}$}
\STATE {$h = g^{\odot_{d'} sk} \in \mathcal{P}_{d', q}$}
\STATE {$pk = (p, d', g, h)$}
\RETURN {$pk, sk$}
\end{algorithmic}
\end{minipage}  
\begin{minipage}[t]{.48\textwidth}
$\texttt{Encrypt}(\textit{msg}, pk)$:

\begin{algorithmic}[1]
\REQUIRE {$\textit{msg} \leq (q - 1)^2$}
\STATE {$(x, y) \leftarrow \textit{msg}$}
\STATE {$d = \frac{x^2 - 1}{y^2} \in \mathbb{F}_q$ with $\chi(d) = - 1$}
\STATE {$m = \frac{x + 1}{y} \in \mathcal{P}_{d, q}$}
\STATE {$r \leftarrow_\$ \{2, \ldots, q\}$}
\STATE {$s = \sqrt{d / d'} \in \mathbb{F}_q$}
\STATE {$c_1 = (g s)^{\odot_d r} \in \mathcal{P}_{d, q}$}
\STATE {$c_2 = (h s)^{\odot_d r} \odot \textit{m} \in \mathcal{P}_{d, q}$}
\RETURN {$c_1, c_2, d$}
\end{algorithmic}

$\texttt{Decrypt}(c_1, c_2, d, pk, sk)$:

\begin{algorithmic}[1]
\STATE {$m = (- c_1^{\odot_d sk}) \odot c_2$}
\STATE {$\textit{msg} \leftarrow \left( \frac{m^2 + d}{m^2 - d}, \frac{2 m}{m^2 - d} \right)$}
\RETURN {\textit{msg}}
\end{algorithmic}
\end{minipage}
\end{center}
\caption{Alternative ElGamal PKE scheme with the cyclic group $(\mathcal{P}_{d, q}, \odot_d)$ of order $q + 1$ and $d$ part of the ciphertext.
Exponentiations are realized with the modified More's algorithm.}
\label{fig:elg2}
\end{figure}

\section{Numerical results}
\label{sec:res}

In this section some practical results about data size and computational costs are presented.
In tables \ref{tab:data} and \ref{tab:perf}, the column ``Points" refers to the first introduced cryptosystem (\cref{fig:brah}), i.e. the ElGamal scheme with the cyclic group $(\mathcal{C}_d(\mathbb{F}_q), \otimes)$ of the points of the Pell hyperbola, which is equivalent to the implementation with a cyclic subgroup of order $q + 1$ of $\mathbb{F}_{q^2}$, because they are isomorphic.
The column ``Parameters" refers to the cryptosystem in \cref{fig:elg}, that is the ElGamal scheme with the cyclic group $(\mathcal{P}_{d, q}, \odot)$ obtained from the parameterization of the Pell hyperbola. The implementation uses the efficient modified More's algorithm for evaluating the exponentiations. Finally, the column ``Alternative" refers to the last introduced cryptosystem (\cref{fig:elg2}) that still works with the cyclic group $(\mathcal{P}_{d, q}, \odot)$ but has $d$ in the ciphertext.

\cref{tab:data} contains the theoretical sizes of the public and secret data involved in the introduced cryptosystems.
In the public key, the parameters of the system are also considered, despite they can be determined only in the first run and then kept public.
Their sizes are the first term of the sum in the first row.
Clearly, since the first cryptosystem works with points on the conic, it has a public key larger than those of the other two cryptosystems.
The public key size in the column ``Alternative" is slightly better than the one for ``Parameters", since $d'$ in the key generation is taken as the smallest non-square.

\begin{table}[t]
\centering
\begin{tabular}{c|ccc}
Data & Points & Parameters & Alternative \\ \hline
Public Key & $4 n + 2 n$ & $3 n + n$ & $2 n + n$ \\
Secret Key & $n$ & $n$ & $n$ \\
Plaintext & $n$ & $n$ & $2 n$ \\
Ciphertext & $4 n$ & $2 n$ & $ 3 n$
\end{tabular}
\caption{Data size in function of the bit length $n$ of $q$ for the ElGamal scheme with the Pell conic points (equivalent to $\mathbb{F}_{q^2}$), with the parameters, and for the alternative ElGamal scheme with $d$ in the ciphertext.}
\label{tab:data}
\end{table}

The second row describes the size of the secret key, which is the same in each column since $1 < sk < q + 1$ in all the cryptosystems.

The maximum length of the plaintext, in row 3, does not include the negligible restrictions required in the first and third cryptosystem in order to succeed in the message encoding into a point of a Pell hyperbola.
It is noteworthy that the last cryptosystem has twice the maximum length than the first two.
This means that, if a message $2n$ bits long needs to be encrypted, the data sizes in the first two columns need to be doubled, while all the values in the third column remain unchanged.
In particular, the ciphertext would be $8n$ bits long for the ElGamal scheme on the ``Points", $4n$ bits long for the one on the ``Parameters" and only $3n$ bits long for the ``Alternative" ElGamal scheme.

\begin{table}[t]
\centering
\begin{tabular}{c|c|ccc}
$n$ & Algorithm & Points & Parameters & Alternative \\ \hline
128 & \texttt{KeyGen} & 0.001428 & 0.000902 & 0.000928 \\
& steps 1-3 & 0.001242 & 0.000599 & 0.000647 \\
& steps 4-7 & 0.000186 & 0.000303 & 0.000281 \\
& \texttt{Encrypt} & 0.000609 & 0.000635 & 0.000895 \\
& \texttt{Decrypt} & 0.000189 & 0.000324 & 0.000350 \\ \hline
256 & \texttt{KeyGen} & 0.003016 & 0.002907 & 0.003306 \\
& steps 1-3 & 0.002329 & 0.002009 & 0.002489 \\
& steps 4-7 & 0.000687 & 0.000898 & 0.000817 \\
& \texttt{Encrypt} & 0.002013 & 0.001883 & 0.002615 \\
& \texttt{Decrypt} & 0.000689 & 0.000962 & 0.001033 \\ \hline
512 & \texttt{KeyGen} & 0.018812 & 0.015633 & 0.007725 \\
& steps 1-3 & 0.015269 & 0.012014 & 0.004717 \\
& steps 4-7 & 0.003543 & 0.003619 & 0.003008 \\
& \texttt{Encrypt} & 0.009528 & 0.007478 & 0.010521 \\
& \texttt{Decrypt} & 0.003580 & 0.003820 & 0.003910 \\ \hline
1024 & \texttt{KeyGen} & 0.073645 & 0.058839 & 0.044672 \\
& steps 1-3 & 0.051974 & 0.039609 & 0.029410 \\
& steps 4-7 & 0.021671 & 0.019230 & 0.015262 \\
& \texttt{Encrypt} & 0.055655 & 0.038504 & 0.056372 \\
& \texttt{Decrypt} & 0.021685 & 0.019608 & 0.019572 \\ \hline
2048 & \texttt{KeyGen} & 0.682581 & 0.317692 & 0.305323 \\
& steps 1-3 & 0.537115 & 0.197664 & 0.213241 \\
& steps 4-7 & 0.145466 & 0.120028 & 0.092082 \\
& \texttt{Encrypt} & 0.366740 & 0.240008 & 0.358340 \\
& \texttt{Decrypt} & 0.146178 & 0.120503 & 0.121528 \\ 
\end{tabular}
\caption{Performance in seconds depending on the bit length $n$ of $q = p$ prime for the ElGamal scheme with the conic points ($\mathbb{F}_{p^2}$), with the parameters and the modified More's algorithm, and for the alternative ElGamal scheme with $d$ in the ciphertext.}
\label{tab:perf}
\end{table}

The second study concerns the efficiency of the three cryptosystems and it is carried out by collecting the elapsed times of a simple implementation in Python of each of the algorithms run on the cluster of the DISMA at Politecnico of Turin.
The case with $q = p$ prime is considered, and the bit length $n$ of $p$ goes from 128 to 2048 by doubling.
The times in the table are the means of 10 randomly generated instances for each size and cryptosystem.
The times for key generation are also divided into the generation of the public parameters (steps 1-3) and the actual public and secret key generation (steps 4-7).
The ElGamal scheme on the ``Points" appears to be slightly better than the other two cryptosystems (in Encryption and Decryption) for small value of $n$, whereas for $n = 512$, $n = 1024$ and $n = 2048$ the performances of the schemes ``Parameters" and ``Alternative" are usually better.
Its key generation is computationally heavier than the others because a valid generator of the conic must be found.
It is important to observe that the ElGamal scheme on the ``Parameters" and the ``Alternative" version have comparable times but, as observed before, the alternative one allows to encrypt messages twice larger than the ones of the other cryptosystems.
Thus, when the message is $2n$ bits long, in order to compare them equally, the times in the third column can be kept unchanged while those for encryption and decryption in the first two columns should be doubled.
For instance, when considering $n = 2048$, if we consider a plaintext of $2n$ bits, the encryption and decryption with the ``Alternative" method takes $0.358340$ and $0.121528$, respectively. The encryption and decryption of the same plaintexts with the ``Parameters" method takes $0.480016$ and $0.241006$, while the ``Points" method (we recall it is like the classical ElGamal over $F_{q^2}$) takes $0.73348$ and $0.292356$, respectively. We would like to highlight that the scheme with the ``Alternative" method allows to decrease the computational costs, since, considering the same size of the plaintetxs, the classical ElGamal scheme needs to be used twice with respect to the scheme based on the ``Alternative" method.

\end{document}